# Charge ordering, stripes and phase separation in manganese perovskite oxides: an STM/STS study

Ch. Renner and G. Aeppli

London Centre for Nanotechnology and Department of Physics and Astronomy, University College London, Gower Street, London WC1E 6BT, UK

H.M. Ronnow[*]

NEC Laboratories America, 4 Independence Way, Princeton, NJ 08540 and James Franck Institute and Department of Physics, University of Chicago, Chicago, IL 60637

[*]Present address -ETH-Zürich & Paul Scherrer Institute, Villigen, Switzerland

*Keywords:* Scanning tunneling microscopy, Strongly correlated electrons, Manganites, Charge ordering, Stripes.

## *Abstract*

A microscopic characterisation of the phase transitions associated with colossal magnetoresistance (CMR) in manganese perovskite oxides is a very important ingredient in the quest of understanding its underlying mechanism. Scanning tunneling microscopy (STM) is most suitable to investigate some of their reported hallmarks, including charge ordering, lattice distortions, and electronic phase separation. Here we investigate $Bi_{1-x}Ca_xMnO_3$ (BCMO) with x=0.76. At this composition, BCMO develops an insulating charge-ordered phase upon cooling, whose study as a function of temperature will allow identifying atomic scale characteristics of the metal-insulator phase transition (MIT). We observe distinct atomic scale phases at temperatures above and below the MIT, with very different electronic and structural characteristics. Combining STM micrographs and current-voltage tunneling characteristics, we find that charge ordering correlates both with the local conduction state (metallic or insulating) and the local structural order. Furthermore, STM shows coexistence of these phases as expected for a first order phase transition.

Colossal magnetoresistance (CMR) – i.e. the large change in resistivity in an applied magnetic field – is an intriguing and potentially useful property. Its occurrence in manganese perovskite oxides is the focus of intense research efforts aiming at understanding the underlying fundamental physics and exploitation in novel devices. The largest magnitude CMR comes alongside a metal-insulator phase transition (MIT), and a detailed microscopic knowledge of the associated metallic and insulating phases is a key ingredient for modelling this phenomenon. A number of characteristic properties of the insulating phase should be readily observable using a scanning tunneling microscope (STM). In particular, the electronic degrees of freedom (charge, spin, and orbital state) are expected to order upon cooling through the MIT [1]. This ordering affects the valence electrons, which are precisely those measured in an STM experiment. Charge ordering (CO) is accompanied by local lattice distortions [2], whose amplitude should be detectable in STM micrographs [3]. Finally, another much-debated hallmark of CMR and the MIT is phase separation into metallic and insulating regions [4]. Although it may seem awkward to study the MIT using STM, which is inoperative on insulators, some CMR manganites remain sufficiently conducting (($\rho \leq 1\Omega$cm) for STM experiments to be carried out in the temperature range of interest.

We have studied single crystals of $Bi_{1-X}Ca_XMnO_3$ (BCMO) using a variable temperature STM (5K – 420K) in ultra-high vacuum. The single crystals were grown using a self-flux technique at a nominal doping of x=0.76. For trivalent Bi and divalent Ca, the Mn ions are in a mixed valence state $Mn^{3+x}$. At high temperature, $Mn^{3+}$ and $Mn^{4+}$ randomly occupy the manganese sites. Upon reducing the temperature, these cations are believed to order, yielding an increased lattice periodicity visible to X-ray and neutron diffraction [3,5]. In the samples we investigated, this occurs at $T_{CO}$=250K, as established using SQUID magnetometry. We used electro-chemically etched tungsten tips, which were subsequently cleaned in-situ by field evaporation against a gold sample. Typical STM set point parameters were U=0.7 V and I=0.2 nA, where the bias voltage is applied to the sample. All the STM micrographs presented below were taken in the constant current mode.

BCMO does not have a natural cleaving plane. Hence it is very difficult to prepare atomically clean surfaces needed for STM investigations, and atomic resolution is difficult to achieve [6,7]. We tried several preparation procedures, including $HNO_3$ etching and in-situ annealing, but all failed to yield atomic resolution micrographs. Only as-grown surfaces cleaned in ultrasound using an organic solvent enabled us to achieve reproducible atomic scale imaging. This cleaning does not remove all surface contamination, and atomic resolution is only possible on a limited number of mostly small terraces (Figure 1). However, whenever stable tunneling conditions with reproducible atomic resolution and spectroscopy were achieved, the resulting micrographs and tunneling current–voltage I(U) characteristics were always the same. Combined with the fact that these results are quantitatively consistent with bulk probes such as X-ray diffraction and optical conductivity, this gives us strong confidence that we are probing intrinsic BCMO properties.

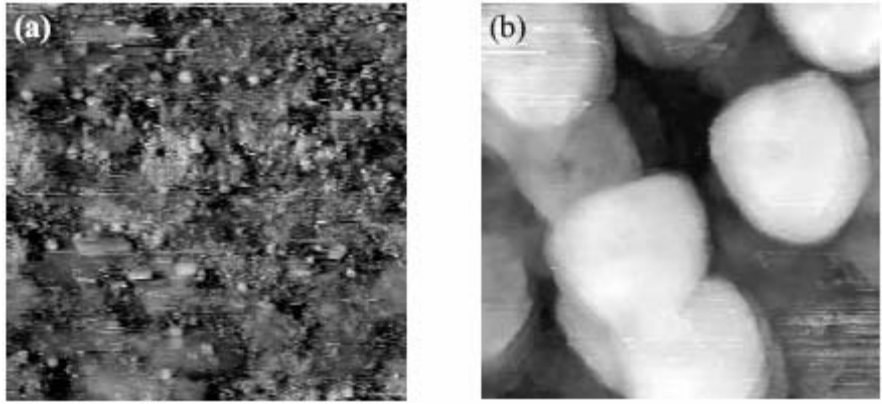

**Figure 1.** Two representative 101 x 101 nm$^2$ STM micrographs of as grown BCMO single crystal surfaces prepared by ultra sound cleaning in an organic solvent. **(a) more typical surface with atomic resolution limited to few small clean areas of about 10x10 nm$^2$. (b) a few surfaces were obtained with significantly larger clean grains. In both cases, we obtain identical micrographs and spectra.**

From these atomically resolved areas, we first focus on the STM micrographs and vacuum tunneling spectra obtained above the MIT, in the metallic paramagnetic phase. The micrographs reveal a square lattice (Figure 2a) consistent with the Mn ion lattice. The average lattice constant $a_0$=3.8±0.1Å is in good agreement with the value of 3.77Å determined by X-ray diffraction [3,8]. The tunneling I(U) characteristics obtained in regions where the above square lattice is resolved show a metallic behaviour with a finite slope at low bias (Figure 3a).

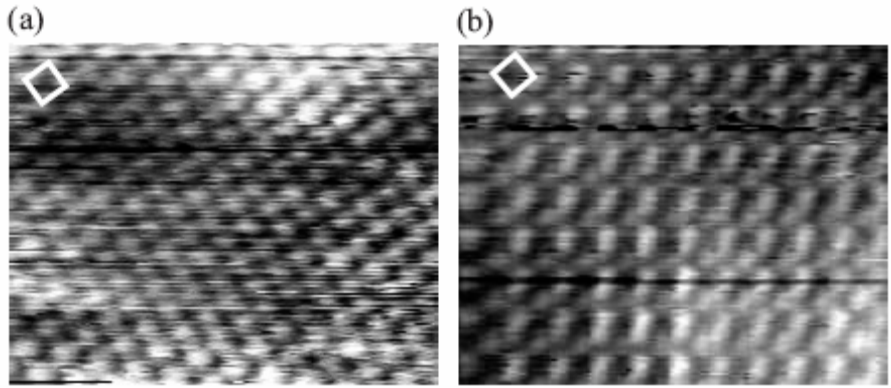

**Figure 2** 5.5 x 4.4 nm$^2$ STM micrographs of (a) the paramagnetic metallic phase at 299 Kelvin and (b) the insulating charge-ordered phase at 100 Kelvin. The micrographs were taken in two separate experiments, hence the slight misalignment of the Mn unit cell depicted as white squares.

We now turn to the STM micrographs and vacuum tunneling spectra obtained below the MIT, in the charge-ordered (CO) phase. Both topography and spectroscopy are dramatically different from their room temperature counterparts. The micrographs show a reconstructed lattice with a $\sqrt{2}a_0 \times \sqrt{2}a_0$ unit cell. The spectroscopy reveals a well-developed gap, of the order of 0.7 eV (Figure 3b). The opening of a gap at the Fermi energy is in agreement with resistivity and optical conductivity measurements [9], both showing the system to become insulating below the CO phase transition temperature $T_{CO}$. The gap amplitude is consistent with low temperature optical conductivity [9], which shows a kink near 0.75 eV.

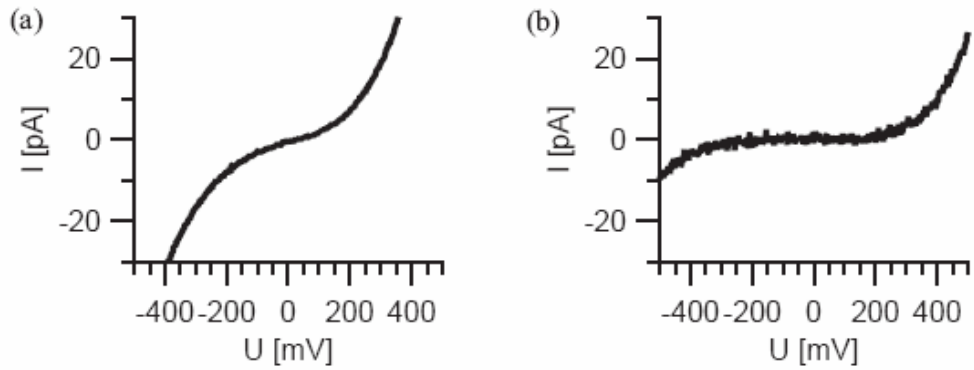

**Figure 3** STM vacuum tunneling spectra of (a) the paramagnetic metallic phase at 299 Kelvin and (b) the insulating charge-ordered phase at 146 Kelvin.

The STM micrographs allow us to visualize the real space electronic and structural features of insulating BCMO with unprecedented atomic scale resolution. While the high temperature micrographs show a rather homogeneous atomic scale map of the surface (detailed analysis reveals some degree of structural disorder [7]), the micrographs of the low temperature CO insulating phase reveal very distinct features. In particular, substantial distortions develop both in amplitude and position of the atoms (Figure 2b). The images can be analysed and compared quantitatively with X-ray structural refinements [3]. This analysis shows that the lateral displacement of the atomic sites can be ascribed to the tilting of the $MnO_6$ Jahn-Teller distorted octahedra. On the other hand, the corresponding distortions alone are too small to account for the enhanced amplitude contrast, which is a consequence of atomic scale charge ordering [7].

The alert reader will have noticed an apparent discrepancy between the checkerboard CO and the nominal hole doping of the BCMO single crystals. Indeed, the observed checkerboard reconstruction corresponds to x=0.5, inconsistent with a nominal Ca content of x=0.76. According to theoretical calculations [10], the checkerboard CO is energetically slightly more favourable than stripe-type CO, which has another degree of freedom (the stripe periodicity) to allow variable valence. Checkerboard regions could be stabilised by defects or surface effects, by intrinsic phase separation, or – less probably – by non-integral valences at the atomic level. Upon close inspection of all our STM data, we found regions at the sample surface where the charges appear to arrange to form stripe-like patterns consistent with the nominal $Mn^{3+}/Mn^{4+}$ ratio of our samples (Figure 4a). The striped CO configuration is only observed at low temperature in the insulating phase. We have not seen it coexisting with the square lattice in the room temperature paramagnetic metallic phase.

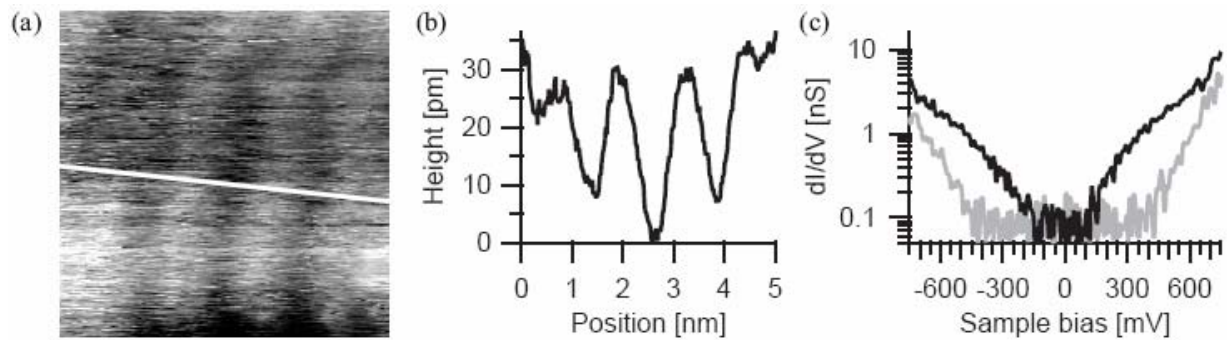

**Figure 4** (a) 5.0 x 5.0 $nm^2$ STM micrograph of charge-ordered stripes in BCMO at 145 Kelvin. (b) Height profile measured along the white line in a. (c) Differential tunneling conductance on a logarithmic scale. The black spectrum measured in the striped region, shows a ~0.3 eV gap at the Fermi energy. For comparison, the grey spectrum measured in a checkerboard charge-ordered region shows a ~0.7 eV gap.

The STM systematically fails to reveal a clear atomic lattice in the striped regions. We merely resolve an amplitude modulation of the order of 0.03nm with a periodicity of about 1.1 nm (Figure 4a,b). Single stripes with a Ca concentration of x=0.76 are expected to have a periodicity $a = 2\sqrt{2}a_0$, which amounts to 1.1nm in close agreement with the STM micrographs. At this stage we can only speculate about the absence of atomic resolution in striped regions. It may result from the mixed valence of the manganese ions, the quasi one dimensionality of the stripes, or a combination thereof. It may also be a direct consequence of atomic scale dynamic fluctuations within the stripes, preventing the STM from resolving individual atoms. Further investigation of this issue is clearly needed. STM spectroscopy in striped regions measures a gap of the order of 0.3 eV at the Fermi energy (Figure 4c), significantly smaller than the gap associated with the checkerboard charge configuration. Interestingly, the gap we measure in the striped regions can account for the low energy tail in the optical conductivity [9] between 0.25 eV and the checkerboard gap of 0.7 eV. This suggests that these two phases do not only appear at the surface, but also in the bulk of the samples.

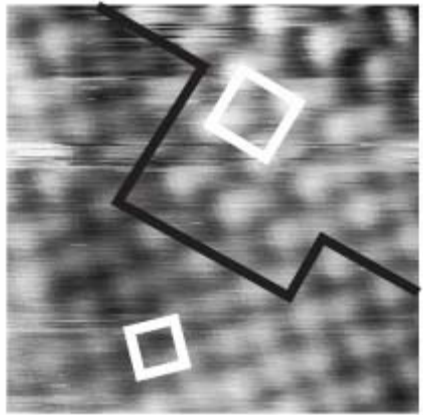

**Figure 5** 3.5 x 3.5 nm² **STM micrograph of the boundary region between a square lattice (bottom left) and the checkerboard charge-ordered phase (top right) in BCMO at 299 Kelvin. The white squares depict the respective unit cells. From [7].**

Finally, we want to address the topic of mesoscopic electronic phase separation, which is matter of a vivid debate in transition metal perovskite oxides. It is being discussed both in the context of colossal magnetoresistance [4,11] and high temperature superconductivity [12]. Provided the insulating phase does not prevent its operation, STM is ideally suited to probe the existence of intermixed insulating and metallic mesoscopic domains in real space. Previously published STM investigations of manganite thin films [13,14] demonstrated phase separation into metallic and insulating regions on sub-micron, but not atomic lengths scales. The main difficulty with these thin films is that the surfaces are very rough on the length scale of typical STM experiments, and atomic resolution has not been achieved so far. On the BCMO single crystals, we identified three distinct phases with high resolution, namely the square lattice and two CO phases (checkerboard and stripes). Below the MIT, we predominantly observe the checkerboard charge ordering. The striped configuration is seen occasionally during the same experiments on the same sample. But the two CO phases always appear in different regions, and we have not been able to observe a boundary region where the two of them join together. Of the 50-100 instances of atomic resolution below the MIT, not a single one showed the metallic square lattice state. Above the MIT, the prevailing atomic scale feature is the square lattice. While we have never observed stripes, we have seen the checkerboard CO phase at room temperature. Moreover, we could image a boundary region between the checkerboard and the square lattice with atomic scale resolution (Figure 5).

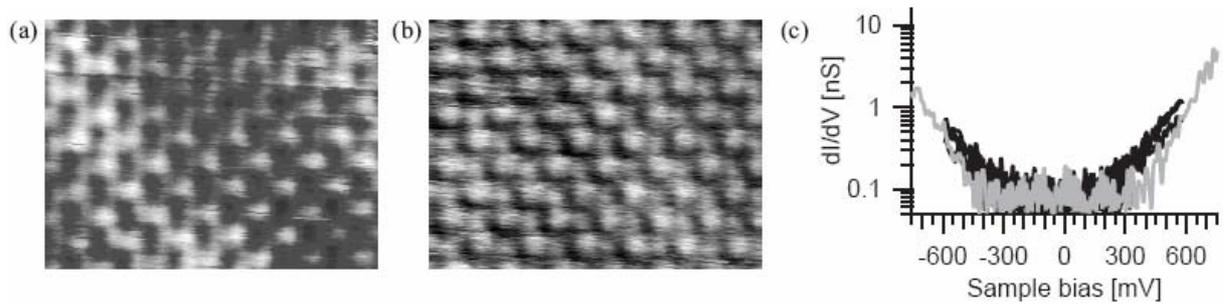

**Figure 6** 4.5 x 3.5 nm$^2$ STM micrographs of BCMO showing (a) the checkerboard charge-ordered phase observed above and (b) below the MIT at 299 Kelvin and 146 Kelvin, respectively. (c) Differential tunneling conductance spectra measured at 299 Kelvin in region a (grey curve) and at 146 Kelvin in region b (black curve). They are shown on a logarithmic scale and normalized to the 299 Kelvin junction resistance *R=U/I* at *U*=0.8V. From [7].

A definitive conclusion about electronic phase separation would require more systematic temperature dependent studies, atomic resolution over larger regions, and better statistics on the relative occurrences. But the fact that we do not observe the square lattice below the MIT suggests that the intrinsic phase separation scenario is not realized in these samples. A scenario better supported by the data we present here, is that the checkerboard phase observed above the MIT is a manifestation of the first order of the charge ordering phase transition, where regions of the insulating phase nucleate already above the transition temperature, possibly stabilised and/or pinned by irregularities of the sample surface. Indeed, the checkerboard CO phase observed at room temperature is exactly the same as the one below the MIT; both have the same microscopic and spectroscopic signatures as illustrated in Figure 6.

In summary, we have presented atomic resolution STM micrographs and vacuum tunneling spectroscopy of the surface of a pseudo-cubic manganese perovskite single crystal above and below the charge-ordering metal-insulator phase transition. The STM micrographs highlight the importance of considering charge ordering along with substantial lattice distortion when developing a model of the insulating phase. Our experiments do reveal coexistence of distinct phases with different atomic and electronic structures, but point towards an extrinsic stabilisation of domains rather than intrinsic phase separation. Tunneling spectra on one hand serve to test microscopic pictures of the different phases, and on the other hand provide valuable input for attempts to model macroscopic properties as a function of temperature and doping.

We thank S-W. Cheong and B-G. Kim for growing high-quality BCMO single crystals. HMR thanks Prof. T. Rosenbaum of the University of Chicago for support and encouragement, and also acknowledges the hospitality of the London Centre for Nanotechnology where this work was completed.

(email: c.renner@ucl.ac.uk)